\documentclass[letterpaper]{jpconf}
\usepackage{graphicx,color}
\usepackage{epsfig,color}
\usepackage{epstopdf}
\usepackage{natbib}
\newcommand{\be}{\begin{eqnarray}}
\newcommand{\ee}{\end{eqnarray}}

\newcommand\gcm{g~cm$^{-3}$}
\newcommand\simgreater{\,\lower0.7ex\hbox{$\stackrel{>}{\sim}$}\,}
\newcommand\simless{\,\lower0.7ex\hbox{$\stackrel{<}{\sim}$}\,}
\newcommand\msol{$M_\odot$}

\newcommand{\nue}{\ensuremath{\nu_{{\rm e}}}}
\newcommand{\nuebar}{\ensuremath{\bar{\nu}_{{\rm e}}}}

\newcommand{\nux}{\ensuremath{\nu_{{\rm x}}}}
\newcommand{\nuxbar}{\ensuremath{\bar{\nu}_{{\rm x}}}}
\newcommand{\num}{\ensuremath{\nu_{\mu}}}
\newcommand{\numbar}{\ensuremath{\bar{\nu}_{\mu}}}
\newcommand{\nut}{\ensuremath{\nu_{\tau}}}
\newcommand{\nutbar}{\ensuremath{\bar{\nu}_{\tau}}}

\begin{document}

\title{Modeling core collapse supernovae in 2 and 3 dimensions with spectral neutrino transport}

\author{S. W. Bruenn$^1$, C. J. Dirk$^1$, A. Mezzacappa$^2$, J. C. Hayes$^3$, J. M. Blondin$^4$, W. R. Hix$^2$ and O. E. B. Messer$^5$}

\address{$^1$ Physics Department, Florida Atlantic University, 777 W. Glades Road, Boca Raton, FL 33431-0991}

\address{$^2$ Physics Division, Oak Ridge National Laboratory, Oak Ridge, TN 37831--6354}

\address{$^3$ Center for Astrophysics and Space Sciences, University of California, San Diego, CA 92093}

\address{$^4$ Department of Physics, North Carolina State University, Raleigh, NC 27695-8202}

\address{$^5$ Center for Computational Sciences, Oak Ridge National Laboratory, Oak Ridge, TN 37831--6354}

\begin{abstract}
The overwhelming evidence that the core collapse supernova mechanism is inherently multidimensional, the complexity of the physical processes involved, and the increasing evidence from simulations that the explosion is marginal presents great computational challenges for the realistic modeling of this event, particularly in 3 spatial dimensions. We have developed a code which is scalable to computations in 3 dimensions which couples PPM Lagrangian with remap hydrodynamics \citep{colellaw84} , multigroup, flux-limited diffusion neutrino transport \citep{bruenn85}, with many improvements), and a nuclear network \citep{hix_t99a}. The neutrino transport is performed in a ray-by-ray plus approximation wherein all the lateral effects of neutrinos are included (e.g., pressure, velocity corrections, advection) except the transport. A moving radial grid option permits the evolution to be carried out from initial core collapse with only modest demands on the number of radial zones. The inner part of the core is evolved after collapse along with the rest of the core and mantle by subcycling the lateral evolution near the center as demanded by the small Courant times. We present results of 2-D simulations of a symmetric and an asymmetric collapse of both a 15 and an 11  {\msol} progenitor. In each of these simulations we have discovered that once the oxygen rich material reaches the shock there is a synergistic interplay between the reduced ram pressure, the energy released by the burning of the shock heated oxygen rich material, and the neutrino energy deposition which leads to a revival of the shock and an explosion.
\end{abstract}

\section{Introduction}
\label{sec:Intro}

The core collapse supernova mechanism remains an unsolved problem despite four decades of effort to unravel it. Evidence has accumulated suggesting that multidimensional effects play an important and perhaps essential role in the mechanism. On the observational side, spectropolarimetry, the large average pulsar velocities, and the morphology of highly resolved images of SN 1987A all suggest that anisotropy develops very early on in the explosion \citep[e.g., see][for reviews and references]{arnettbkw89, mccray93, nomoto_skys94}. On the theoretical side, analyses of immediate post-bounce core profiles given by computer simulations show that a variety of fluid instabilities are present and may play a role in the explosion mechanism \citep[e.g., see][for a review]{buras_rjk06}. In particular, multi-dimensional numerical simulations have shown that convective overturn in the neutrino-heated region behind the stalled shock may be important for the success of the neutrino-driven mechanism as they help transport hot gas from the neutrino-heating region directly to the shock, while downflows simultaneously carry cold, accreted matter to the layer of strongest neutrino heating where a part of this gas readily absorbs more energy from the neutrinos. These simulations have also revealed that a non-radial, low-mode standing accretion shock instability (SASI) may also grow, given time, via an advective-acoustic cycle \citep{foglizzo_sj05}, or the propagation of sound waves \citep{blondin_m06}. This low-mode distortion of the shock, whether it arises from the coalescence of higher mode convective eddies or a SASI, may be at the root of some of the above mentioned supernova observables.

\section{A Computational Challenge}

The complexity of the supernova mechanism precludes a purely analytic investigation, requiring, instead  realistic numerical simulations for its unraveling. This presents great technical challenges. A typical supernova explosion energy is $10^{51}$ ergs, or 1B (a unit of energy, the bethe, which honors Hans Bethe, who spent more than a decade contributing to core collapse supernova theory), and must be regarded as marginal, being of the same order as the gravitational binding energy of the envelope of the progenitor ejected. On the other hand, 100 times this energy resides in the internal energy of the immediate post collapse core, and the near negative of this in the form of gravitational binding energy. Thus, simulations must be energy conserving to high accuracy if we are to take their outcomes seriously. Ultimately $\sim$ 300 B in energy is released by the core in neutrinos of all flavors, and their interaction with the stellar core and mantle will either power the explosion itself or play a major role in the explosion dynamics. An inaccurate treatment of neutrino transport can qualitatively change the results of a simulation. Since neutrinos can originate deep within the core where neutrino mean free paths are small compared with other relevant length scales and propagate out to regions where the reverse is true, the transport scheme must be accurate in both regimes plus the all-important intermediate regime where the critical neutrino energy deposition occurs. Neutrinos interact with matter in a variety of energy dependent ways and this demands that both the neutrino transport and the interactions receive a full spectral implementation, rather than having the neutrino spectrum prescribed. The angular distribution of the neutrinos is also important to compute accurately. In particular, it affects the neutrino heating, and the latter is primarily determined in a region where the angular distribution can neither be assumed to be isotropic nor radially free streaming. Supernova simulations must be carried out in two, and preferably three, spatial dimensions for the reasons mentioned . The nuclear abundances should be evolved in regions where nuclear statistical equilibrium (NSE) cannot be maintained. This will enable the potentially observable products of nucleosynthesis to be followed and the energy released by nuclear burning to be fed back into the computation of the explosion dynamics. While the energy released is expected to be rather small, it could be locally significant and have an influence on the dynamics if all other factors give rise to a very marginal outcome. Finally, general relativistic effects must be incorporated as they influence the size of the neutrino heated region, the rate of matter advection through this region, and the neutrino luminosities and {\small RMS} energies \citep{bruenndm01}, and can therefore profoundly affect the dynamics.

\section{Brief Description of the Code}

Here we briefly describe the code we have developed to simulate core collapse supernovae. (A more complete description will be given in \citep{bruenn_dmhnhm06}). The code attempts to incorporate most of the above criteria for realistic core collapse modeling while being efficient enough to evolve progenitors from the onset of collapse to the order of 1 sec post bounce given present day state-of-the-art computational resources, such as the Cray XT3. It conserves total energy (gravitational, internal, kinetic, and neutrino) to within $\pm$ 0.5 B. The code currently has three main components:  a hydro component, a neutrino transport component, and a nuclear reaction network component.

The hydrodynamics is evolved via a Godunov finite-volume scheme, specifically, a Lagrangian remap implementation of the  Piecewise Parabolic Method (PPM) \citep{colellaw84}. Being third order in space (for equal zoning) and second order in time the code is well suited for resolving shocks, composition discontinuities, etc. with modest gridding requirements. The scheme is currently Newtonian but GR corrections at successively more sophisticated levels are being added. A moving radial grid option wherein the radial grid follows the average radial motion of the fluid makes it possible for the core infall phase to be followed with good resolution. The equation of state (EOS) of Lattimer and Swesty \citep{lattimers91} is currently employed for matter in NSE above $1.7 \times 10^{8}$ \gcm. Below this density matter in NSE is described similarly by 4 nuclei (neutrons, protons, helium, and a representative heavy nucleus) in a highly modified version of the EOS described by \citep{cooperstein85a}. For regions not in NSE, an EOS with a nuclear component consisting of 14 alpha particle nuclei ($^{4}$He to $^{60}$Zn), protons, neutrons, and an iron-like nucleus is used. An electron-positron EOS with arbitrary degeneracy and degree of relativity spans the entire density-temperature regime of interest. To avoid the ``odd-even decoupling'' and carbuncle phenomenon for shocks aligned parallel to a coordinate axis we have employed grid jittering in these calculations up to 5 ms post bounce. At this time the grid jittering was turned off and the velocity field seeded with random perturbations at 0.1\% the sonic speed. We feel now that grid jittering is unsatisfactory as it affects the entire flow, and is being replaced by an algorithm that is local to the shocks only. The gravitational potential is solved by means of the Newtonian gravity spectral Poisson solver described in \citep{muller_s95}.

Ideally, neutrino transport should be implemented with full multi-D Boltzmann transport. This important effort is being made \citep{cardall_rem05a} but will be very computationally expensive. We compromise by implementing a ``ray-by-ray-plus'' approximation \citep[cf.][]{buras_rjk06} for neutrino transport whereby the lateral effects of neutrinos such as lateral pressure gradients (in optically thick conditions), neutrino advection, and velocity corrections are taken into account, but transport is performed only in the radial direction. Transport is computed by means of multigroup flux-limited diffusion with a flux limiter that has been tuned to reproduce Boltzmann transport results to within a few percent \citep{liebendorfer_mbmbct04}. All O($v/c$) observer corrections have been included. The transport solver is fully implicit and solves for four neutrino flavors simultaneously (i.e., \nue's, \nuebar's, \num's and \nut's (collectively \nux's), and \numbar's and \nutbar's (collectively \nuxbar's)), allowing for neutrino-neutrino scattering and pair-exchange, and different $\nu$ and $\bar{\nu}$ opacities. The PPM technology has  been directly applied to both the spatial and energy advection of neutrinos in both the radial and lateral directions. The neutrino opacities employed for the simulations are the ``standard'' ones described in \citep{bruenn85} with the isoenergetic scattering of nucleons replaced by the more exact formalism of \citep{reddypl98} which includes nucleon blocking, recoil, and relativistic effects, and with the addition of nucleon-nucleon bremsstrahlung \citep{hannestadr98} with the kernel reduced by a factor of five in accordance with the results of \citep{hanhart_pr01}.

The nuclear composition in the non-NSE regions of these models is evolved by the thermonuclear reaction network of \citep{hix_t99a}.  This is a fully implicit general purpose reaction network, however in these models only reactions linking the 14 alpha nuclei from $^{4}$He to $^{60}$Zn are used. Data for these reactions is drawn from the REACLIB compilations \citep{rauscher_t00}. The nucleons have only very small abundances at any time and are included to make the NSE-nonNSE transition smoother. The iron-like nucleus is included to conserve charge in a freezeout occurring with an electron fraction below 0.5 \citep[cf.][]{kifonidisojm03} The advection of material across an NSE - nonNSE interface in either direction is performed as detailed in \citep{bruenn_dmhnhm06}. Also, entire zones are moved from NSE to nonNSE as conditions dictate.

\section{Progenitors}

We follow the evolution from infall to $\sim$0.7 sec. post bounce of an 11 \msol\ progenitor, S11s7b, and a 15 \msol\ progenitor, S15s7b, both provided by \citep{woosley95}. Each progenitor is evolved twice; once from the progenitor configuration as given (symmetric infall), and once with a 5 percent density increase in a conical wedge of angular width 22.5 degrees centered about the polar axis (asymmetric infall). A minimal resolution of 192 radial and 32 angular zones was used. As a preliminary test of the resolution dependence of these simulations the evolution of S11s7b with symmetric infall was repeated with the angular resolution increased  to 96 angular zones. Finally, we restarted the simulation initiated from S11s7b with the nuclear  nuclear reaction network turned off. A summary of the simulations with descriptive nomenclature is given below. The radial grid was programmed to follow the mean radial motion of the fluid until 1 ms post bounce at which time the radial grid was frozen.
\bigskip

\begin{center}
\begin{tabular}{|l|c|c|l|} \hline
\multicolumn{4}{|c|}{\bfseries Simulations} \\ \hline
\itshape Model & \itshape Progenitor & \itshape Number of Rays & \itshape Infall \\ \hline
11M\_Sym\_32R & 11\msol S11s7b & 32 & symmetric \\ \hline
11M\_Sym\_32R\_nonuc & 11\msol S11s7b & 32 & symmetric \\ \hline
11M\_Sym\_96R & 11\msol S11s7b & 96 & symmetric \\ \hline
11M\_Asym\_32R & 11\msol S11s7b & 32 & asymmetric \\ \hline
15M\_Sym\_32R & 15\msol S15s7b & 32 & symmetric \\ \hline
15M\_Asym\_32R & 15\msol S15s7b & 32 & asymmetric \\ \hline
\end{tabular}
\end{center}
\bigskip

\section{Results}

\subsection{Below the Neutrinosphere}

A recent study of protoneutron star convection with many references is given by \citep{dessart_blo06}. We note a few features of our simulations and postpone a more detailed discussion to \citep{bruenn_dmhnhm06}. Convective-like fluid motions in the nascent neutron stars of our models are always subsonic and confined to a layer between 10 and 25 km. Gravity waves are also evident and grow to modest amplitudes during the course of a simulation. We do not see the large-scale low-mode gravity waves reported by \citep{burrows_ldom06}, but it is unclear at this time whether our spherical grid is in some way suppressing these modes.

\subsection{Above the Neutrinosphere}

\begin{figure}[!h]
\vspace{-3.cm}
\hspace{2.cm}
\setlength{\unitlength}{1.0cm}
{\includegraphics[width=11.0 cm]{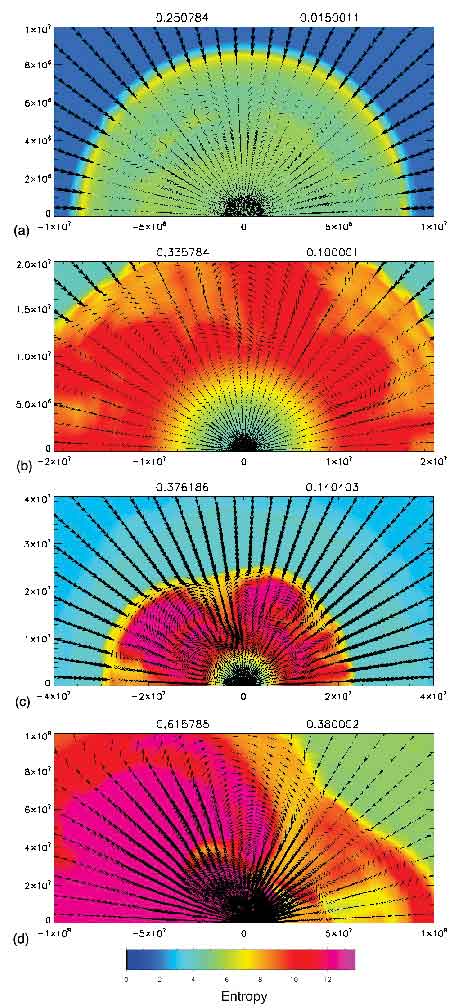}}
%{\includegraphics[width=\columnwidth]{11M_evolve.eps}}
\caption{\label{11M_evolve}
Entropy and velocity configuration snapshots of the Model 11M\_Sym\_32R.}
\end{figure}

\begin{figure}[!h]
\vspace{-3.cm}
\hspace{2.cm}
\setlength{\unitlength}{1.0cm}
{\includegraphics[width=11.0 cm]{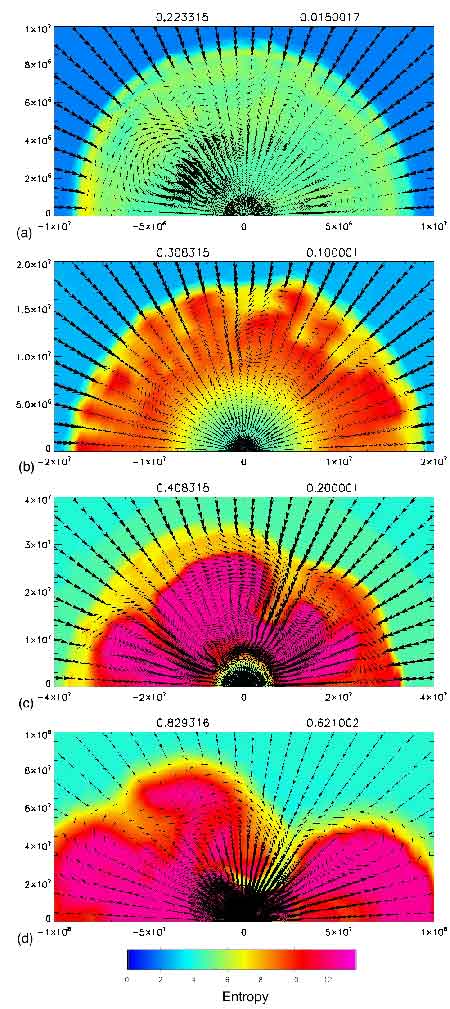}}
%{\includegraphics[width=\columnwidth]{15M_evolve.eps}}
\caption{\label{15M_evolve}
Entropy and velocity configuration snapshots of the Model 15M\_Sym\_32R.}
\end{figure}

Within 10 - 20 ms after bounce and following the initial propagation of the shock, low contrast entropy plumes appear and extend out to about 80 km. This is illustrated for models 11M\_Sym\_32R and 15M\_Sym\_32R in Figures \ref{11M_evolve}a and \ref{15M_evolve}a, respectively. These arise from the Ledoux-unstable negative entropy gradient left over by the weakening shock \citep{arnett86, burrowsf92}, extending from about 40 to 80 km. They gradually mix and merge and become indistinct after $\sim$ 40 ms. A global angular mass average of the entropy at this time shows that its radial profile has been flattened by this overturn and mixing. The negative initial radial gradient of the angular mass average of the electron fraction is reduced by not eliminated. 

More dramatic is the flow that develops afterwards in the region between the gain-radius and the shock. (The gain--radius is the boundary separating the region suffering net neutrino cooling below from net neutrino heating above. By $\sim$50 ms post bounce neutrino heating establishes within this region a negative entropy gradient thereby rendering this region Ledoux-unstable. Distinct high-entropy plumes begin to be seen at this time separated by narrower low-entropy downfows. The latter originate from a Rayleigh-Taylor unstable layer of lower entropy material newly accreted through the shock. This pattern becomes very pronounced by 100 ms (see Figures \ref{11M_evolve}b and \ref{15M_evolve}b) by which time the shock has been pushed out quasistatically to $\sim$250 km for model 11M\_Sym\_32R and 190 km for 15M\_Sym\_32R by the rapid accretion of material through the shock. The rising high entropy plumes begin to push the shock outward causing local dome-like distortions. This flow pattern has been observed in other simulations \citep[e.g.][]{kifonidisojm03,scheck_kjm06,buras_rjk06a,burrows_ldom06}. The high entropy plumes merge and grow, distorting the shock even more. By 140 ms for model 11M\_Sym\_32R (Figure \ref{11M_evolve}c) and 200 ms for model 15M\_Sym\_32R (\ref{15M_evolve}c) two or three plumes dominate the flow. The shock, now showing large dome-like distortions, deflects the incident matter flux when the latter is not normal to the shock interface. This is particularly evident in the vicinity of the shock depressions, where the incoming flow is funneled into low entropy downflows. At this time the shock begins to exhibit global distortions of a quasi-oscillatory time-dependent character, which we provisionally attribute to either the advective-acoustic cycle or the SASI. These global distortions exhibit a tendency to be most pronounced along the polar directions. Our simulations are performed on a polar grid and assume axisymmetry; the polar axis is therefore impenetrable for the fluid flow. Therefore, converging flows are directed along the polar axis, either inward or outward. Additionally, 2D axisymmetry on a polar grid leads to a zoning in which the equatorial tori are much larger than those near the poles. Ideally, 3D simulations should be performed on a grid without coordinate singularities.

\subsection{Interaction between the Shock and the Oxygen Layers}

Most interesting, and seemingly unreported before, is the behavior of the shock once it has been penetrated by the $^{16}$O layer. Initially, $^{16}$O at the 0.1 mass fraction level (0.5 mass fraction level)  is at 1400 km (1600 km) for model 11M\_Sym\_32R, and at 2500 km (6400 km) for model 15M\_Sym\_32R. For 11M\_Sym\_32R a combination of the inward advection of the $^{16}$O layer and the bipolar oscillation of the shock results in $^{16}$O at the 0.1 mass fraction level penetrating the shock along the positive polar axis at $\sim$160 ms post bounce. For model 15M\_Sym\_32R this does not occur until $\sim$450 ms post bounce. A synergistic interplay between neutrino heating, the reduced ram pressure of the less dense $^{16}$O-rich material, and the energy released by the $^{16}$O burning sets in at this time. The reduced ram pressure of the $^{16}$O-rich material causes the shock to readjust to a larger radius. At the same time the energy released by the burning of the shock heated oxygen, though small (\simless 0.1 B) (but not locally small), increases the pressure behind the shock and causes it to move further out. This, in turn, increases the important ratio of the advective to the neutrino heating time scale \citep[see][]{buras_rjk06a} to values exceeding unity and the shock begins to power up. This stage is illustrated in Figures \ref{11M_evolve}d and \ref{15M_evolve}d for models 11M\_Sym\_32R and 15M\_Sym\_32R, respectively. For model 11M\_Sym\_32R the shock expansion is almost unipolar, with a strong downflow becoming established near the polar axis. For 15M\_Sym\_32R, the shock expansion is more dipolar with a strong downflow being established closer to the equator. A unipolar versus a bipolar shock expansion will result in a larger neutron star kick, and the emergence of one or the other may be largely stochastic and result in a bimodal pulsar velocity distribution, as pointed out by \citep{scheck_kjm06}. The estimated explosion energies at the time the simulations were terminated were rather weak, $\sim$1 B for model 11M\_Sym\_32R and $\sim$.3 B 15M\_Sym\_32R, but not enough time has elapsed for the explosions to become fully developed. The protoneutron star rest masses were 1.42 \msol and 1.54 \msol for models 11M\_Sym\_32R and 15M\_Sym\_32R, respectively.

\begin{figure}[!h]
\vspace{-3.cm}
\hspace{2.cm}
\setlength{\unitlength}{1.0cm}
{\includegraphics[width=12.0 cm]{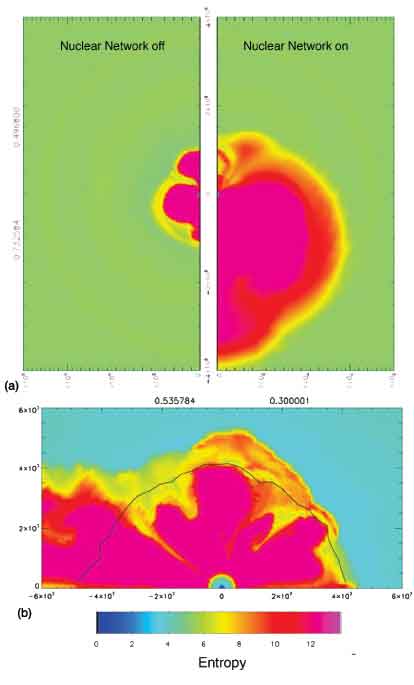}}
\caption{\label{Nuclear}
(a) Comparison of models 11M\_Sym\_32R\_nonuc (left) and 11M\_Sym\_32R (right) at 500 ms post bounce. (b) the boundary separating matter in NSE (inner) from matter not in NSE (outer) for model 11M\_Sym\_96R as the explosion is powering up. The matter not in NSE consists at this time of oxygen rich material.}
\end{figure}

To test how much of the shock revival seen in these models when the oxygen layer penetrated the shock was a function of the reduced ram pressure versus the energy released by the burning of the shocked oxygen rich material, we restarted model 11M\_Sym\_32R at 130 ms post bounce with the nuclear network turned off (model 11M\_Sym\_32R\_nonuc). The difference between the two simulations at 500 ms post bounce is shown in Figure \ref{Nuclear}a. At this time model 11M\_Sym\_32R is undergoing an explosion whereas model 11M\_Sym\_32R\_nonuc is still exhibiting bipolar oscillations with no clear sign of an explosion developing. Clearly the energy released by nuclear burning is an essential component of the neutrino energy deposition, reduced ram pressure, nuclear burning trinity for these simulations. Figure \ref{Nuclear}b shows for  model 11M\_Sym\_96R the penetration through the shock of the oxygen-rich material as the explosion is powering up at 300 ms post bounce.

\subsection{Symmetric versus Asymmetric Infall}

Surprisingly, not much difference was noted in the long-term evolution of the models depending on whether the simulations were initiated from a symmetric or asymmetric collapse. After some initial wobbling of the core the evolution settled down largely to the scenario described above for the symmetric model. If anything, the evolution of 15M\_Asym\_32R proceeded for the first 400 ms post bounce with less bipolar shock distortions than model 15M\_Sym\_32R. Eventually model 11M\_Asym\_32R underwent a unipolar explosion very similar to model 11M\_Sym\_32R, while model 15M\_Asym\_32R also underwent a unipolar explosion, unlike its counterpart 15M\_Sym\_32R. It appears that the SASI depends on a long shock revival delay for its development, and is less sensitive to the initial particulars of the shock.

\subsection{Dependence on Angular Resolution}

The overall evolution of model 11M\_Sym\_96R is similar to that of model 11M\_Sym\_32R although some of the details are different. Features, particularly downflows, are narrower. The explosion powers up slightly earlier for model 11M\_Sym\_32R, but the global flow is quite different for the two models, being more bipolar for the 11M\_Sym\_96R. These differences in the global flow patterns may simply be a statistical phenomenon; more simulations will be needed to settle this issue.

\subsection{Conclusions}

We have developed a numerical code coupling multi-dimensional hydrodynamics, a nuclear reaction network, and spectral neutrino transport in a ``ray-by-ray-plus'' approximation to simulate core collapse supernova from the infall epoch to $\sim$1 sec post bounce. We have performed simulations initiated from an 11\msol\ and a 15\msol\ progenitor. Two simulations were initiated from each progenitor, one with a spherically symmetric infall and one with an infall made asymmetric by a slight polar overdensity. Neutrino driven convection is observed for all models beginning $\sim$50 ms post bounce. High entropy plumes separated by lower entropy downflows are evident by 100 ms post bounce for all models. The plumes begin to merge so that by 200 ms two or three large bubbles remain and begin to distort the shock. Bipolar oscillations of the shock begin to become evident at this time as well, which we attribute to a SASI. Most interesting is the synergistic interplay between the reduced ram pressure, the energy released by the burning of the shocked oxygen-rich material, and neutrino heating that ensues once the oxygen rich layers penetrate the shock. All three ingredients appear to be essential, and result in the shock being pushed out into the unburnt material and the powerup of an explosion.

The results reported here are very promising in that many of the supernova observables may be reproduced, but they need to be viewed with caution. The simulations need to repeated with GR incorporated into the code, and in 3-dimensions, preferably with the use of a singularity free grid.

\ack

The authors would like to acknowledge partial funded by a grant from the DOE Office of Science Scientific Discovery through Advanced Computing Program.

\def\newblock{\hskip .11em plus .33em minus .07em}
\bibliography{BibTeX_list}

\begin{thebibliography}{10}

\bibitem{colellaw84}
P.~Colella and P.~R. Woodward.
\newblock {\em Journal of Computational Physics}, 54:174--201, 1984.

\bibitem{bruenn85}
S.~W. Bruenn.
\newblock {\em Astrophysical Journal Supplement}, 58(4):771--841, August 1985.

\bibitem{hix_t99a}
W.~R. Hix and F.-K. Thielemann.
\newblock {\em Journal of Computational and Applied Mathematics}, 109:321--351,
  September 1999.

\bibitem{arnettbkw89}
W.~D. Arnett, J.~N. Bahcall, R.~P. Kirshner, and S.~E. Woosley.
\newblock {\em Annual Review of Astronomy and Astrophysics}, 27:701--756, 1987.

\bibitem{mccray93}
R.~McCray.
\newblock {\em Annual Review of Astronomy and Astrophysics}, 31:175--216, 1993.

\bibitem{nomoto_skys94}
K.~Nomoto, T.~Shigeyama, S.~Kumagai, H.~Yamaoka, and T.~Suzuki.
\newblock Amsterdam: Elsevier/North-Holland, 1994.

\bibitem{buras_rjk06}
R.~Buras, M.~Rampp, H.~Th. Janka, and K.~Kifonidis.
\newblock {\em Astronomy and Astrophysics}, 447:1049--1092, March 2006.

\bibitem{foglizzo_sj05}
T.~Foglizzo, L.~Scheck, and H.-T. Janka.
\newblock {\em Astronomy and Astrophysics, submitted, astro-ph/0507636}, 2005.

\bibitem{blondin_m06}
J.~M. Blondin and A.~Mezzacappa.
\newblock {\em Astrophysical Journal}, 642:401--409, May 2005.

\bibitem{bruenndm01}
S.~W. Bruenn, K.~R. DeNisco, and A.~Mezzacappa.
\newblock {\em Astrophysical Journal,}, 560:326--338, October 2001.

\bibitem{bruenn_dmhnhm06}
S.~W. Bruenn, C.~J. Dirk, A.~Mezzacappa, J.~C. Hayes, J.~M. Blondin, W.~R. Hix,
  and O.~E.~B. Messer.
\newblock {\em in preparation}, 2006.

\bibitem{lattimers91}
J.~M. Lattimer and F.~D. Swesty.
\newblock {\em Nuclear Physics A}, 535:331--376, December 1991.

\bibitem{cooperstein85a}
J.~Cooperstein.
\newblock {\em Nuclear Physics A}, 438:722--739, 1985.

\bibitem{muller_s95}
E.~M\"{u}ller and S.~Steinmetz.
\newblock {\em Computer Physics Communications}, 89:45--58, August 1995.

\bibitem{cardall_rem05a}
C.~Y. Cardall, A.~O. Rasoumov, E.~Endeve, and T.~Mezzacappa.
\newblock In {\em Proceedings of Open Issues in Understanding Core Collapse
  Supernovae, National Institute for Nuclear Theory, University of Washington,
  22-24 June 2004}. World Scientific, in press, and astro-ph/0510704, 2005.

\bibitem{liebendorfer_mbmbct04}
M.~Liebend\"{o}rfer, O.~E.~B. Messer, A.~Mezzacappa, S.~W. Bruenn, C.~Y.
  Cardall, and F.-K. Thielemann.
\newblock {\em Astrophysical Journal Supplement}, 150:263--316, January 2004.

\bibitem{reddypl98}
S.~Reddy, M.~Prakash, and J.~M. Lattimer.
\newblock {\em Physical Review D}, 58:013009--1, May 1998.

\bibitem{hannestadr98}
S.~Hannestad and G.~Raffelt.
\newblock {\em Astrophysical Journal}, 507:339--352, November 1998.

\bibitem{hanhart_pr01}
C.~Hanhart, D.~R. Phillips, and S.~Reddy.
\newblock {\em Physics Letters B}, 499:9--15, February 2001.

\bibitem{rauscher_t00}
T.~Rauscher and F.-K. Thielemann.
\newblock Astrophysical reaction rates from statistical model calculations.
\newblock {\em Atomic Data and Nuclear Data Tables}, 75:1--351, May 2000.

\bibitem{kifonidisojm03}
K.~Kifonidis, T.~Plewa, H.-Th. Janka, and E.~M\"{u}ller.
\newblock {\em Astronomy and Astrophysics}, 408:621--649, September 2003.

\bibitem{woosley95}
S.~E. Woosley.
\newblock Private Communication, 1995.

\bibitem{dessart_blo06}
L.~Dessart, A.~Burrows, E.~Livne, and C.~D. Ott.
\newblock {\em Astrophysical Journal, in press; astro-ph/0510229}, 2006.

\bibitem{burrows_ldom06}
A.~Burrows, E.~Livne, L.~Dessart, C.~D. Ott, and J.~Murphy.
\newblock {\em Astrophysical Journal}, 640:878--890, April 2006.

\bibitem{arnett86}
W.~D. Arnett.
\newblock volume IAU Symposium 125. Dordrect: Reidel), 1986.

\bibitem{burrowsf92}
A.~Burrows and Fryxell.
\newblock {\em Science}, 258:430--434, October 1992.

\bibitem{scheck_kjm06}
L.~Scheck, K.~Kifonidis, H.~Th. Janka, and E.~Mueller.
\newblock {\em Astronomy and Astrophysics, in press; astro-ph/0601302}, 2006.

\bibitem{buras_rjk06a}
R.~Buras, M.~Rampp, H.~Th. Janka, and K.~Kifonidis.
\newblock {\em Astronomy and Astrophysics, submitted; astro-ph/0512189}, 2006.

\end{thebibliography}

\end{document}